\documentclass[preprint,12pt]{elsarticle}

\usepackage{amssymb}
\usepackage{amsmath}
\usepackage{float}

\journal{Journal of Non-Crystalline Solids}
\usepackage{booktabs}
\begin{document}

\begin{frontmatter}



\title{Atomistic insights into the structural, thermal, and mechanical evolution of $Zr_{47.5}Cu_{47.5}Ag_{5}$ bulk metallic glass.}
\author[inst1]{Mendez, N.}
\author[inst2]{Jaroszewicz, S.}
\author[inst3]{Beccar-Varela, M. P.}
\author[inst3]{Mariani, M. C.}

\affiliation[inst1]{organization={Instituto Sabato, UNSAM, CNEA. },
            addressline={Av.Gral Paz 1499}, 
            city={Buenos Aires},
            postcode={C1650}, 
            country={Argentina}}
            
\affiliation[inst2]{organization={Departamento de Materia Condensada, CNEA},
            addressline={Av.Gral Paz 1499}, 
            city={Buenos Aires},
            postcode={C1650}, 
            country={Argentina}}
            
\affiliation[inst2]{organization={Department of Mathematical Sciences, UTEP},
            addressline={500 W. University Ave.}, 
            city={El Paso},
            postcode={TX 79968},
            country={USA}}

\begin{abstract}
Bulk metallic glasses (BMGs) are distinguished by amorphous atomic structures that confer superior mechanical performance; however, the evolution of these properties in ternary bulk configurations remains insufficiently explored. In this study, we employed large-scale molecular dynamics simulations to investigate the structural, thermal, and mechanical properties of $Zr_{47.5}Cu_{47.5}Ag_{5}$ BMGs. Our thermodynamic and topological analyses, utilizing potential energy regression and the Modified Wendt-Abraham parameter, identified a glass transition temperature ($T_g$) of approximately $692\text{ K}$. Structural characterization via Voronoi tessellation and partial radial distribution functions reveals that the amorphous matrix is stabilized by icosahedral clusters, with Ag atoms inducing significant chemical heterogeneity through localized nano-clustering. Thermal transport properties, computed via the Green-Kubo formalism, demonstrate a monotonic decrease in conductivity with temperature, consistent with structural scattering saturation in disordered lattices. Mechanical tensile testing reveals that the material exhibits robust rate- and temperature-dependent behavior, with yield strengths reaching $\approx 2.3\text{ GPa}$ at room temperature. We show that macroscopic strain-softening is intrinsically linked to the thermally induced collapse of rigid icosahedral motifs, which facilitates shear band percolation. These findings provide a structural rationale for the beneficial role of Ag dopants in enhancing the resilience of multicomponent metallic glasses.
\end{abstract}

\begin{highlights}
\item Vitrification of Zr-Cu-Ag metallic glass via MD simulations at $10^{10}$ K/s.
\item Ag-doping promotes nano-clustering and structural disruption in Zr-Cu matrix.
\item Thermal conductivity aligns with the Cahill-Pohl structural scattering limit.
\item Mechanical yield strength and softening correlate with icosahedral collapse.
\end{highlights}


\end{frontmatter}


\section{Introduction}
\label{sec:introduction}

Metallic glasses (MGs) represent a distinct class of materials characterized by a lack of long-range translational symmetry, which confers superior mechanical and physical properties compared to their crystalline counterparts \cite{Wang2009, TREXLER2010759}. Among various alloy systems, Zr--Cu--Ag-based metallic glasses have garnered significant research interest due to their exceptional glass-forming ability (GFA) and structural versatility . While existing literature has extensively characterized these alloys in thin-film architectures \cite{Dassonneville2021, HUSZAR2023139822, JAIN2023170728}, comprehensive insights into their structural, thermal, and mechanical evolution within bulk metallic glass (BMG) configurations remain limited, necessitating systematic atomistic investigations.

\vspace{2mm}

The functional performance of these amorphous alloys is intrinsically linked to their thermodynamic stability \cite{INOUE2000279}. Specifically, the glass transition temperature ($T_g$) serves as a fundamental benchmark for evaluating the structural resilience of MGs under thermal and mechanical loading \cite{kato2008relationship,lu2000correlation,DMOWSKI2007125}. Zr--Cu--Ag systems exhibit high $T_g$ values, which broaden their operational window for high-temperature applications. Below $T_g$, these materials possess high strength and hardness\cite{Zhang01052006}, coupled with a remarkable capacity to resist wear and corrosion \cite{QIN20063713, OAK2009322}, the latter being facilitated by the formation of a self-passivating surface oxide layer. 

\vspace{2mm}

Furthermore, the mechanical behavior of Zr--Cu--Ag glasses is highly sensitive to the proximity of the glass transition \cite{DATYE2020152979}. As the temperature approaches $T_g$, the viscosity of the amorphous matrix changes significantly, dictating the crossover from brittle shear localization to more ductile flow regimes \cite{SPAEPEN1977407,ARGON197947}. Optimizing this relationship is essential to exploit these materials in high-stress environments, such as aerospace, automotive, and biomedical engineering, where thermal stability and mechanical toughness are critical requirements.

\vspace{2mm}

Despite recent progress, the interplay between the composition-dependent $T_g$, local atomic topology, and macroscopic mechanical response remains a subject of intensive study. This paper investigates the structural, thermal, and mechanical properties of $Zr_{47.5}Cu_{47.5}Ag_{5}$ bulk metallic glasses. Through large-scale molecular dynamics simulations, we elucidate how structural heterogeneities and local bonding configurations dictate the glass transition, thermal transport, and deformation mechanisms of this ternary system.

The remainder of this paper is organized as follows: Section \ref{sec:Simulation details} describes the computational methodology and simulation parameters. Section \ref{sec:results} presents the structural, thermal, and mechanical results, followed by a detailed discussion. Finally, Section \ref{sec:conclusion} summarizes the key findings and implications of this work.

\section{Materials and methods}
\subsection{Simulation details}
\label{sec:Simulation details}

Molecular dynamics simulations were conducted using the LAMMPS package \cite{plimpton1995fast} with an integration time step of 1 fs. The $Zr_{47.5}Cu_{47.5}Ag_5$ system, comprising 8192 atoms in a cubic cell with periodic boundary conditions , was modeled using a embedded-atom method (EAM) potential \cite{Chen2011}. To generate the amorphous structure, the system was equilibrated at 2000 K for 1 ns in the NPT ensemble to ensure a homogeneous melt. Subsequently, the sample was quenched to 300 K at a constant cooling rate of $10^{10}$ K/s. The resulting glass was further relaxed at 300 K and 0 Pa for 10 ns to eliminate residual stresses.

All MD simulations were performed on a high-performance computing platform, using a parallel computing approach with 32 CPU cores. The total simulation time for each tensile test was approximately 100 ns. The statistical uncertainties in the calculated mechanical properties were estimated by performing multiple independent simulations with different initial conditions. The uncertainties were found to be less than 5\% for all calculated properties.

\subsection{Structural Analysis Methods}
\label{subsec:glass_transition_methods}
The glass transition temperature ($T_g$) marks the crossover from a metastable supercooled liquid to a solid-like amorphous state. In this work, we determined $T_g$ using two independent thermodynamic and structural indicators to ensure consistency. 

First, we analyzed the temperature dependence of the potential energy per atom ($E_p$). We identified two distinct linear regimes corresponding to the glassy and supercooled liquid states. These regimes were fitted using:

$$E_p(T) = a + bT$$

The value of $T_g$ was defined as the temperature at the intersection of these two regressions. To minimize statistical bias, the fitting ranges were chosen sufficiently far from the transition region (e.g., 400–600 K for the glass and 1000–1200 K for the liquid).

\vspace{2mm}

Additionally, we employed the Modified Wendt-Abraham (MWA) parameter \cite{Wendt1978}, $R$, defined as $R = g_{min} / g_{max}$, where $g_{min}$ and $g_{max}$ are the first minimum and maximum of the radial distribution function, respectively. This structural parameter provides a more sensitive measure of the local atomic ordering during vitrification than just the potential energy.

To complement the thermodynamic analysis and further quantify the local atomic symmetry, we calculated the Bond-Orientational Order (BOO) parameters, $q_l$, as proposed by Steinhardt et al\cite{Steinhardt1983}. Specifically, $q_4$ and $q_6$ were monitored to detect potential crystalline nucleation. These parameters allow for a clear differentiation between disordered glassy environments and common crystalline lattices (e.g., BCC, FCC, and HCP), ensuring the structural integrity of our amorphous samples.

\vspace{2mm}

To characterize the local atomic packing, we performed a Voronoi tessellation analysis. We identified the coordination polyhedra by their Voronoi indices $\langle n_3, n_4, n_5, n_6 \rangle$, where $n_i$ denotes the number of faces with $i$ edges. This topological descriptor allows for the precise classification of icosahedral-like environments ($\langle 0,0,12,0 \rangle$) and larger Kasper polyhedra, which are known to be the fundamental structural motifs in Zr-based metallic glasses \cite{Park2006}.

\vspace{2mm}
To complement the topological descriptors, the local structural entropy was computed following the fingerprinting approach proposed by Piaggi and Parrinello \cite{Piaggi2017}. This method relies on the local environmental density distribution to quantify the degree of order around each atom, providing a robust metric to distinguish between disordered liquid-like configurations and locally ordered motifs. The local entropy descriptor $S_{loc}$ is defined based on the probability distribution of neighbor environments, effectively acting as an order parameter that is sensitive to subtle symmetry breaking within the glassy matrix. This parameter serves as an effective thermodynamic proxy to identify regions of arrested structural dynamics and local chemical heterogeneity, offering a high-resolution map of the vitrification process at the atomic scale.

\vspace{2mm}

All topological and structural analyses were performed utilizing the MDAPy Python package, which provides a comprehensive framework for post-processing molecular dynamics data \cite{mdapy2023}.

\subsection{Thermal Transport Methods}
\label{subsec:thermal_methods}
The equilibrium thermal conductivity ($\kappa$) was determined using the Green-Kubo formalism \cite{GreenKubo1}, which relates the transport coefficient to the fluctuations of the heat flux vector ($J$) in an equilibrium state:
$$\kappa = \frac{V}{3k_B T^2} \int_0^\infty \langle J(t) \cdot J(0) \rangle \, \text{d}t$$
where $V$ is the volume and $T$ the temperature. The heat flux vector was calculated using the atomic velocities and the stress tensor derived from the EAM potential. For each simulation temperature, the system was equilibrated in the NVT ensemble, followed by an NVE production run of 1 ns to compute the heat flux autocorrelation function (HCACF). To ensure robust convergence, the results represent the average of 5 independent simulations, with production windows carefully analyzed to mitigate transient edge effects by extracting data from the plateau region of the autocorrelation function.

\vspace{2mm}

The specific heat capacity at constant volume ($C_v$) was simultaneously computed from the fluctuations of the total energy in the NVT ensemble, ensuring thermodynamic consistency between transport and equilibrium properties. Statistical uncertainty intervals for $C_v$ were estimated as the standard deviation derived from the plateau region of the energy fluctuation series, providing a conservative assessment of the statistical dispersion within the converged regime.

\subsection{Mechanical Characterization via Uni-axial Tension}
\label{subsec:mechanical_methods}
The mechanical response of the BMG samples was investigated through simulated uni-axial tensile loading, implemented in the LAMMPS environment using the strain-controlled method. The deformation was applied by imposing a constant strain rate ($\dot{\epsilon}$) along the x-axis, while maintaining zero pressure in the transverse directions (y and z) via the NPT ensemble to allow for lateral structural relaxation. Tests were conducted at temperatures ranging from 300 K to 900 K with strain rates of $10^8$ and $10^9\text{ s}^{-1}$. Atomic configurations were visualized and analyzed using the OVITO software \cite{ovito}, enabling the tracking of shear transformation zones and void evolution during the deformation process.

\vspace{2mm}

To validate the accuracy of the MD simulations, the calculated mechanical properties were compared with the experimental data available in the literature \cite{Liu2012, Yang2015}. The simulation results were also compared with the predictions of existing theoretical models for BMGs, such as the shear-transformation-zone theory and the free-volume theory \cite{Falk1998, Greer1999}.

\section{Results and discussion}
\label{sec:results}
\subsection{Vitrification and Global Structure}

\begin{figure}[H]
\centering
\includegraphics[width=1\textwidth]{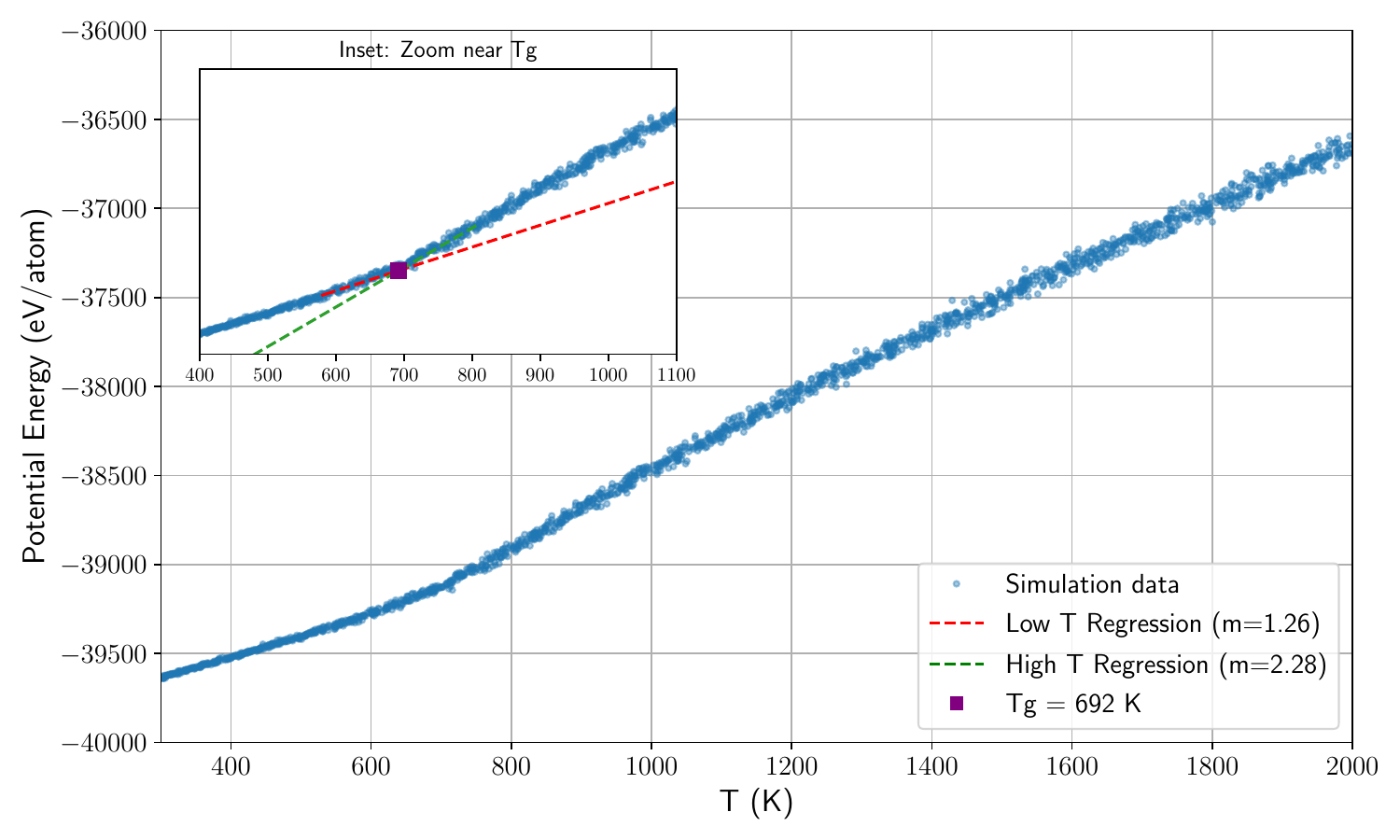}
\caption{\textbf{Temperature dependence of the potential energy during the melt-quenching process for Zr$_{47.5}$Cu$_{47.5}$Ag$_{5}$.} The glass transition temperature ($T_g = 692\text{ K}$) is determined by the intersection of the two linear regression lines representing the liquid and glassy regimes. The inset highlights the distinct change in slope (inflection point) around the structural arrest region.}
\label{fig:energy}
\end{figure}

 The structural and thermodynamic evolution of the Zr$_{47.5}$Cu$_{47.5}$Ag$_{5}$ alloy exhibits two distinct temperature regimes with a continuous, smooth transition between them (Fig. \ref{fig:energy}). To quantify this transition, the potential energy values within both temperature ranges were independently fitted using linear regression functions as mentioned in Sec. \ref{subsec:glass_transition_methods}. The glass transition temperature ($T_g$) was subsequently estimated from the intersection point of these two linear trends (Fig. \ref{fig:energy}, inset). Following this approach, a value of $T_g = 692\text{ K}$ was obtained, establishing a robust thermodynamic baseline for the vitrified state.

\begin{figure}[H]
\centering
\includegraphics[width=1\textwidth]{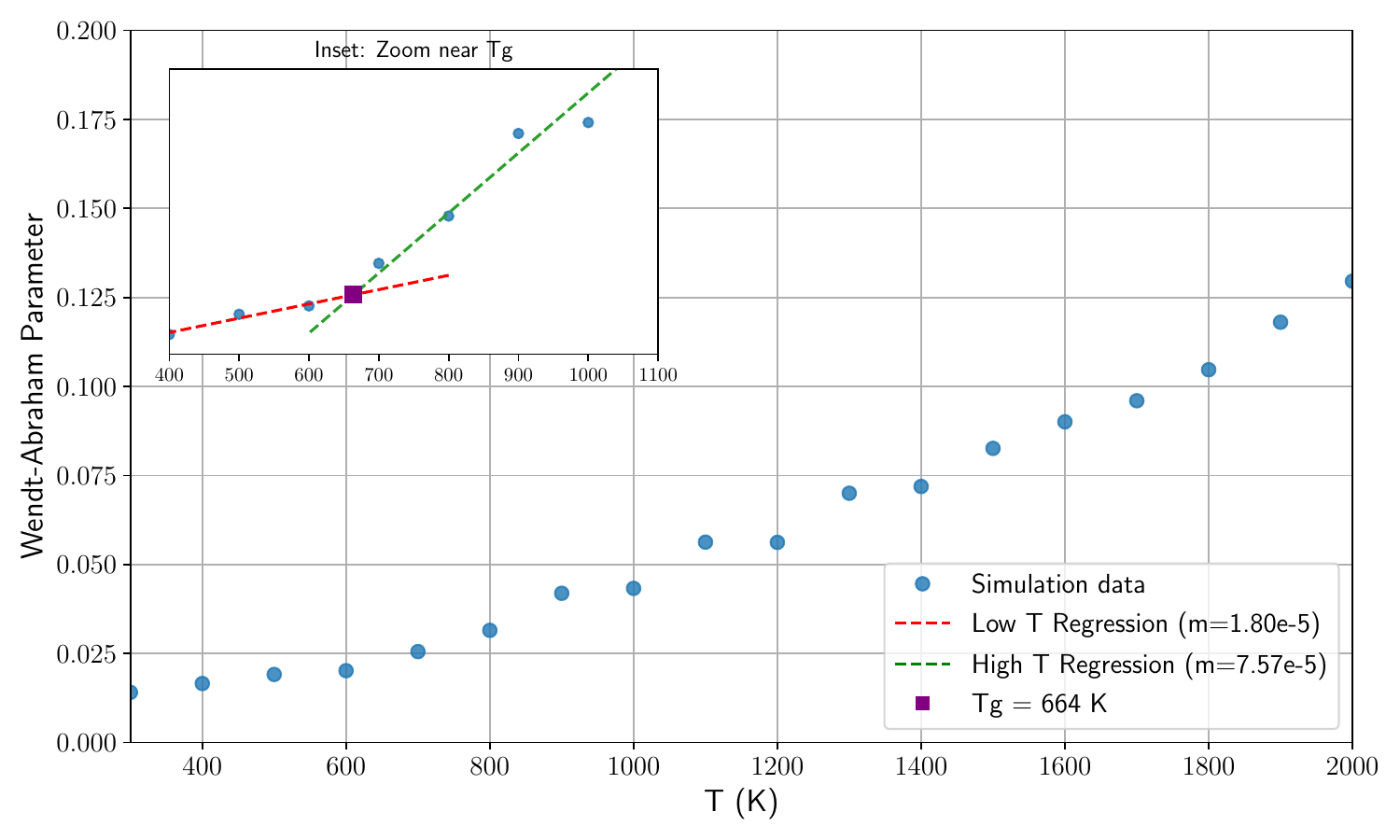}
\caption{\textbf{Temperature dependence of the Modified Wendt-Abraham (MWA) parameter for Zr$_{47.5}$Cu$_{47.5}$Ag$_{5}$.} The structural glass transition temperature ($T_g \approx 664\text{ K}$) is extracted via the intersection of the low- and high-temperature linear regimes. The plot validates the vitrification behavior observed in the potential energy data, capturing the structural arrest from a structural, short-range order perspective.}
\label{fig:mwa}
\end{figure}

To confirm these results the modified Wendt-Abraham parameter (MWA) introduced by Celtek et al. \cite{Celtek} was used to calculate Tg. Figure \ref{fig:mwa} shows the temperature dependence of the MWA parameter for the Zr$_{47.5}$Cu$_{47.5}$Ag$_5$ alloy simulated under 0 GPa pressure. Results for other pressures are not presented to avoid redundancy. The graph displays two different slopes in the high temperature range (800-1100 K) and low temperature range (400-650 K), with linear fits drawn for both regions. $T_g$ was determined to be approximately $664\ K$ by extrapolating these lines to their intersection point.
Notably, the slight discrepancy ($\sim 28\text{ K}$) between the glass transition temperatures obtained from both methods is entirely expected. This variation originates from the fundamental contrast between the potential energy, which acts as a macroscopic thermodynamic indicator of the system's global state, and the MWA parameter, which tracks the vitrification process strictly from a local, topological short-range order perspective.

\begin{figure}[H]
\centering
\includegraphics[width=0.85\textwidth]{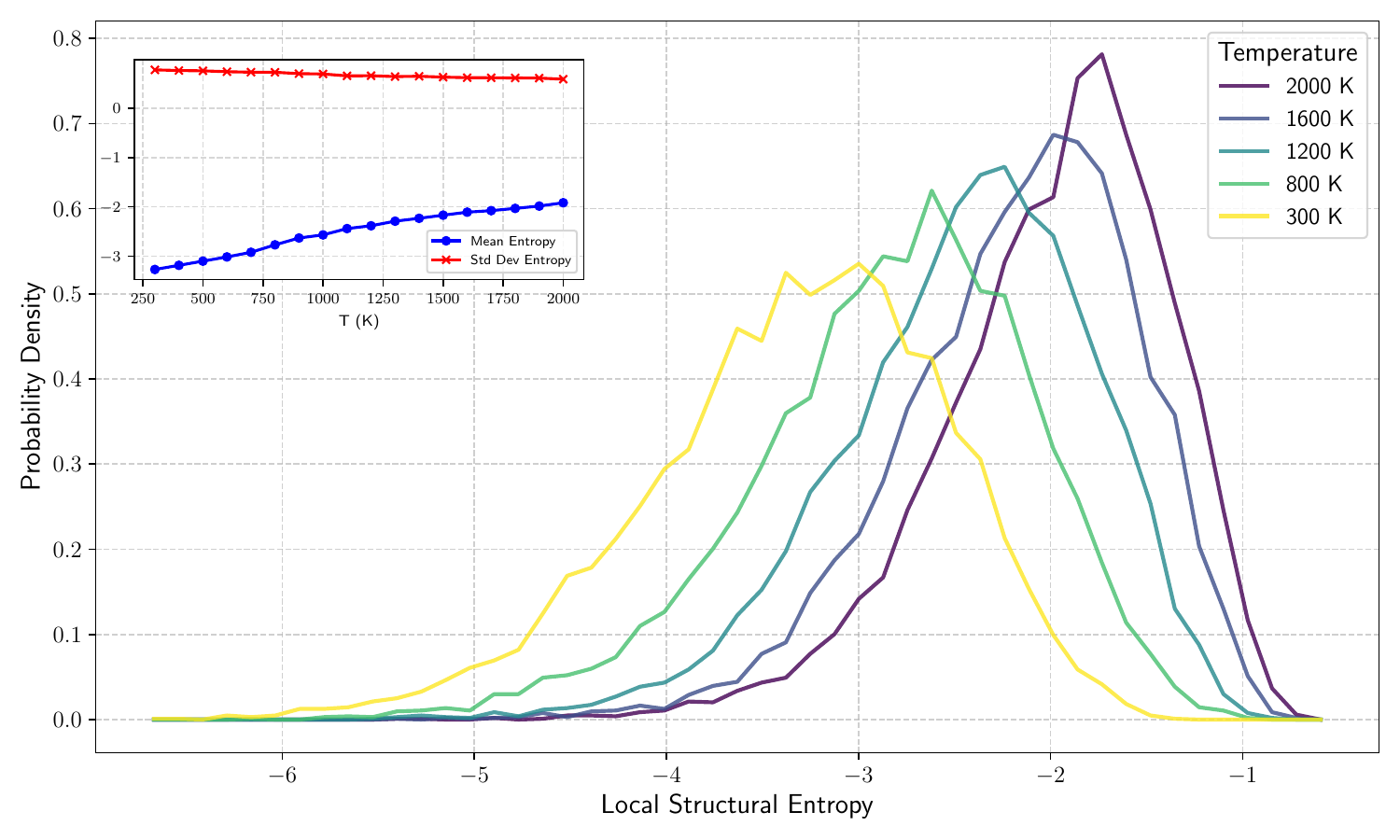}
\caption{\textbf{Temperature dependence of the local configurational entropy ($S_{\text{local}}$).} Probability density functions (PDFs) calculated during the melt-quenching process. The inset displays the evolution of the distribution's statistical parameters (mean $\langle S_{\text{local}} \rangle$ and standard deviation $\sigma$) as a function of temperature, highlighting the structural arrest and local differentiation.}
\label{fig:entropy}
\end{figure}

To complement this thermodynamic and structural description of the vitrification process, the evolution of the local configurational entropy ($S_{\text{local}}$) was investigated. Specifically, the probability density functions (PDFs) of $S_{\text{local}}$ for the entire atomic system were computed as a function of temperature (Fig. \ref{fig:entropy}). As the temperature decreases from the homogeneous melt at 2000 K down to the glassy state at 300 K, the $S_{\text{local}}$ distribution exhibits a monotonic shift toward lower values, representing the progressive loss of accessible configuration space. Interestingly, this shift is accompanied by a very slight widening of the PDF profile at lower temperatures. This behavior reflects that while the global system loses entropy during structural arrest, the local atomic environments undergo significant differentiation, freezing into distinct structural configurations due to the emergence of chemical and topological heterogeneities. Moreover, below the glass transition region ($T_g \approx 692\text{ K}$), a practical independence of the distribution's mean features with temperature is observed. This plateau clearly indicates the complete structural arrest of the constituent atoms, confirming the effective transformation from a metastable supercooled liquid into a structurally stable amorphous matrix.

\begin{figure}[H]
\centering
\includegraphics[width=0.8\textwidth]{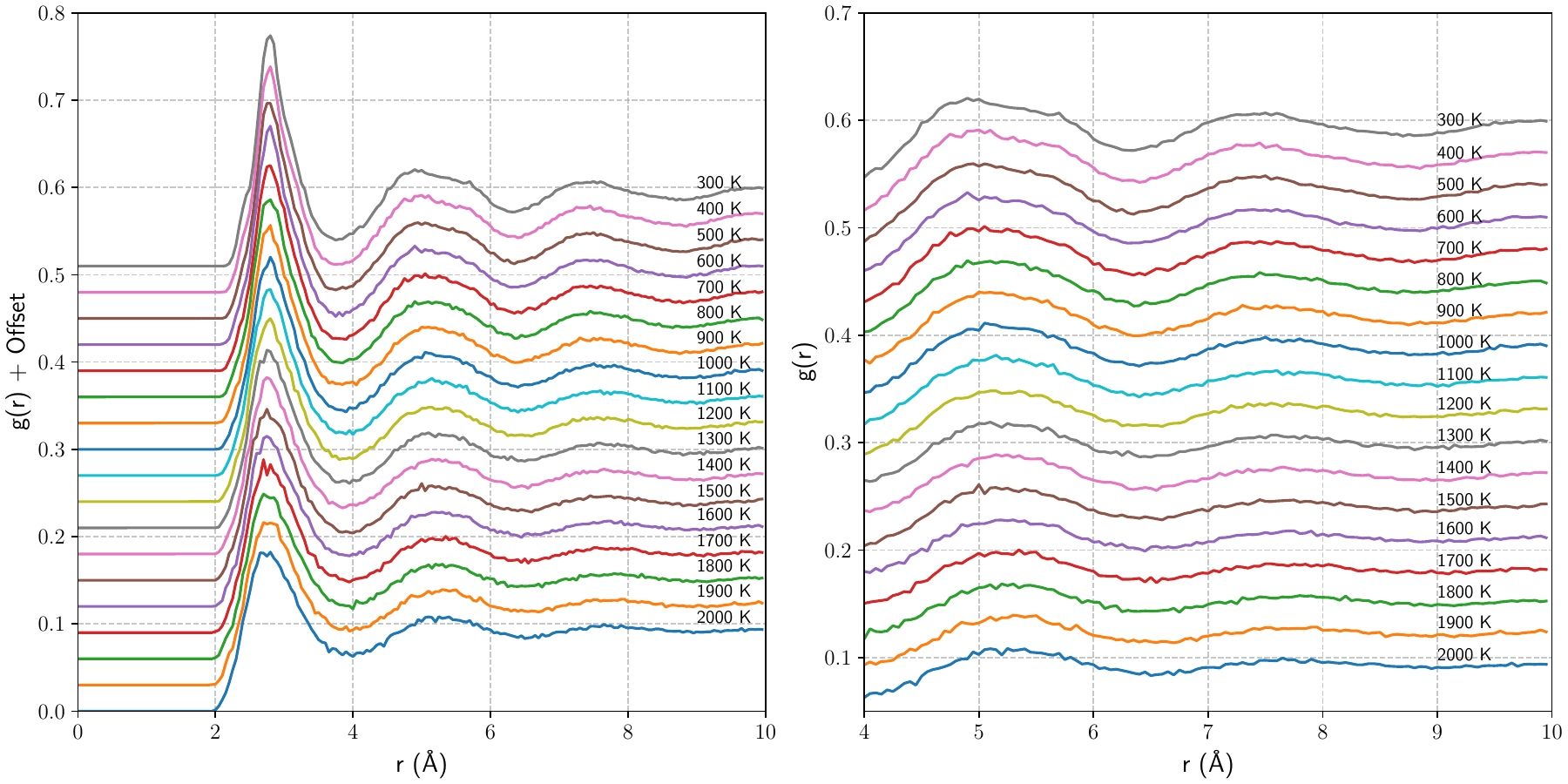}
\caption{\textbf{Total radial distribution functions ($g(r)$) at selected temperatures during quenching.} The right panel highlights the evolution of the second peak, demonstrating the distinctive splitting feature below the glass transition.}
\label{fig:gr}
\end{figure}

\begin{figure}[H]
\centering
\includegraphics[width=0.8\textwidth]{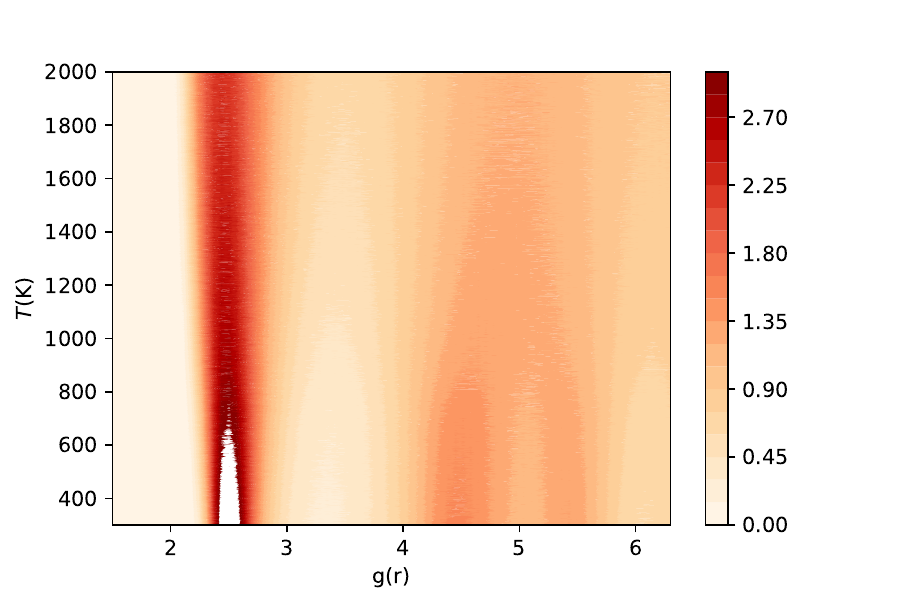}
\caption{\textbf{Contour plot of the total radial distribution function $g(r)$ as a continuous function of temperature.} The sharp emergence of the second-peak splitting around $T_g$ acts as a structural fingerprint of vitrification.}
\label{fig:contour_gr}
\end{figure}

This structural arrest and the local differentiation observed through the configurational entropy are directly reflected in the geometric spatial correlations of the alloy. Consequently, the structural evolution during the cooling process was characterized by the radial distribution function, $g(r)$, as shown in Figure \ref{fig:gr} and \ref{fig:contour_gr}. At high temperatures ($T \geq 1300$ K), the system exhibits a liquid-like profile with broad, diffuse peaks. However, as the temperature decreases below the glass transition region ($T_g \approx 692$ K), a distinct splitting of the second peak emerges. This feature signifies the emergence of short-to-medium range order and the effective freezing of the atomic structure into a glassy state.

\subsection{Topological and Chemical Short-order Range}
\label{sec:structural_properties}
\begin{figure}[H]
    \centering
    \includegraphics[width=0.95\linewidth]{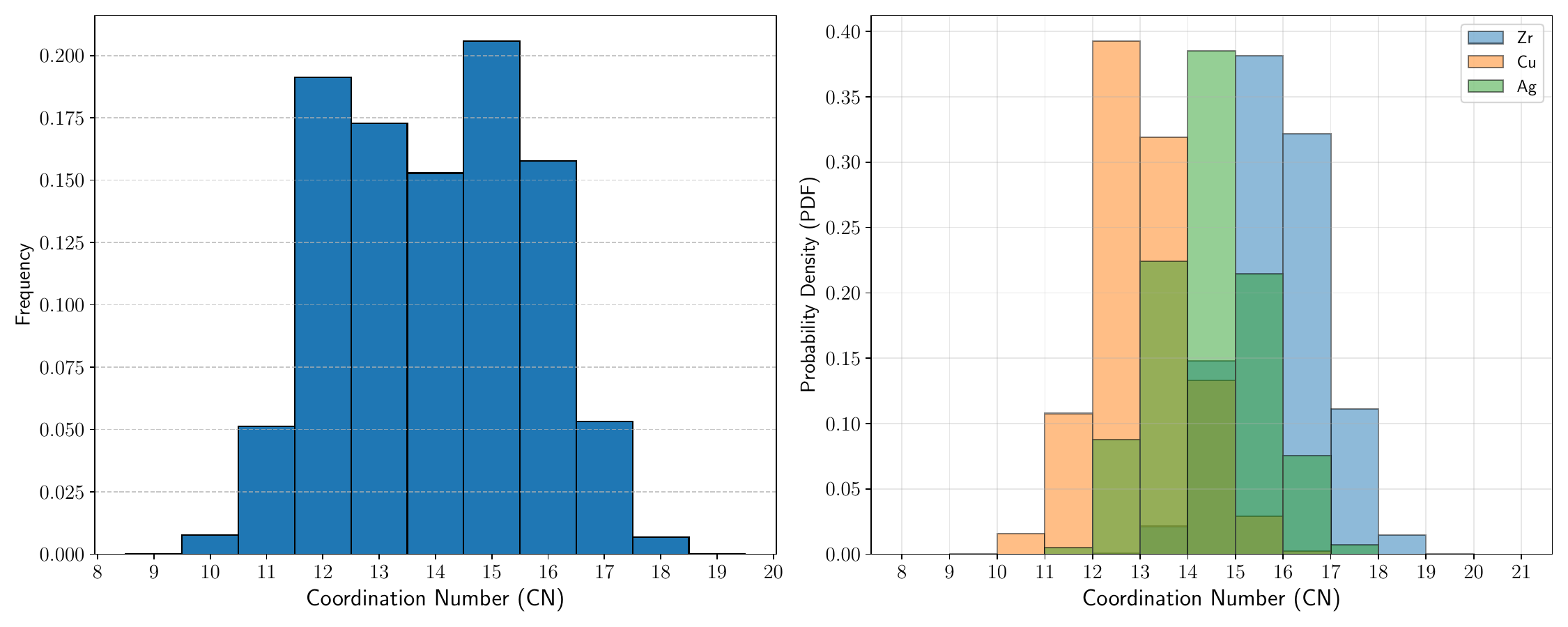}
    \caption{\textbf{Topological short-range order characterization via Voronoi tessellation at 300 K.} (a) Global probability density function (PDF) of coordination numbers (CN) for the Zr$_{47.5}$Cu$_{47.5}$Ag$_{5}$ bulk metallic glass, highlighting a prominent bimodal distribution centered at $CN=12$ and $CN=15$. (b) Species-dependent partial CN distributions. Smaller Cu atoms display a sharp preference for icosahedral-like environments ($CN=12$), whereas the larger Zr atoms dictate higher coordination motifs ($CN=15-16$) to optimize local packing density. Ag atoms exhibit an intermediate behavior centered at $CN=14$, reflecting their distinct role within the amorphous matrix.}
\label{fig:voronoi_cn_combined}
\end{figure}

To unravel the local atomic configurations underlying the macroscopic structural arrest, a comprehensive topological characterization of the vitrified alloy at 300 K was performed via Voronoi tessellation. The global distribution of coordination numbers ($CN$), derived from the polyhedral faces, is presented in Figure \ref{fig:voronoi_cn_combined}a. The distribution spans a broad range from 10 to 18 neighbors, quantitatively confirming the highly disordered nature of the bulk metallic glass and the successful suppression of long-range crystalline order during the rapid quenching process. Notably, the topological profile exhibits a well-defined bimodal character with predominant populations strictly centered at $CN=12$ and $CN=15$. This distinct topology reflects the competitive coexistence of highly packed, icosahedral-like local environments and larger polyhedral clusters typical of the Zr--Cu--Ag ternary system. Such geometric diversity creates a highly frustrated local energy landscape, which acts as the primary structural barrier against crystalline nucleation.

The species-dependent coordination number distributions (Fig. \ref{fig:voronoi_cn_combined}b) clearly demonstrate the topological and chemical short-range order (CSRO) within the vitrified Zr$_{47.5}$Cu$_{47.5}$Ag$_{5}$ alloy. The smaller Cu atoms exhibit a sharp, prominent preference for $CN=12$, which directly aligns with the high population of highly packed icosahedral clusters embedded in the matrix. In contrast, the larger Zr atoms accommodate significantly higher coordination numbers ($CN=15-16$) to satisfy local packing density requirements, effectively acting as a rigid structural backbone that mechanically supports the smaller icosahedral units. Interestingly, the dopant Ag atoms exhibit an intermediate behavior, strictly centered at $CN=14$. This structural behavior aligns with previous findings indicating that minor Ag additions favor glass formation in Zr--Cu systems \cite{Yang2015, Celtek}. Due to their intermediate atomic radius, Ag atoms act as local structural disruptors within the primary Zr--Cu glassy network, successfully modifying the connectivity without triggering lattice ordering. This multi-component atomic size mismatch promotes a frustrated geometric energy landscape, thereby providing a structural rationale for the enhanced robustness of the amorphous matrix against crystalline nucleation.

\begin{figure}[H]
\centering
\includegraphics[width=0.75\textwidth]{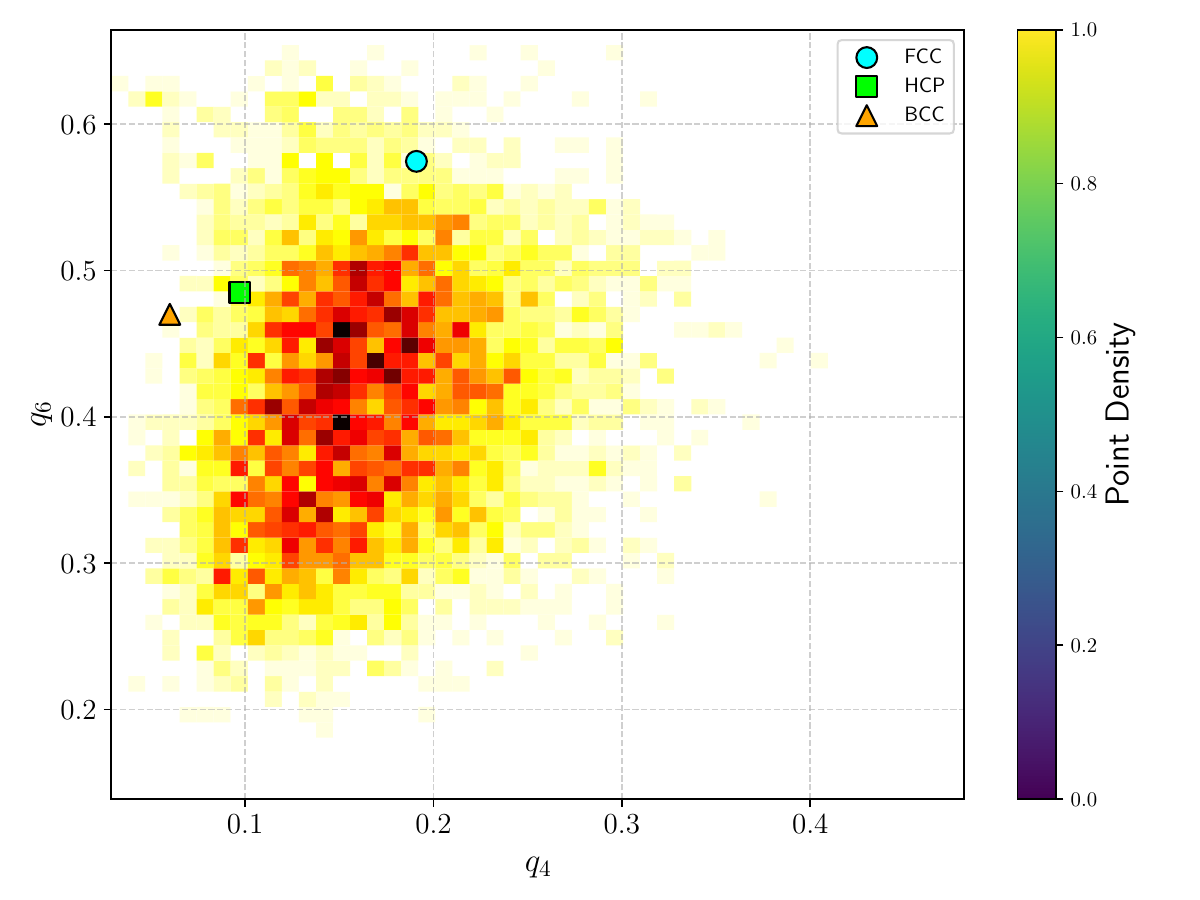}
\caption{\textbf{Assessment of local bond-orientational order via Steinhardt parameters at $T = 300\text{ K}$.} Point density distribution of $q_6$ versus $q_4$ computed for all atoms within the vitrified Zr$_{47.5}$Cu$_{47.5}$Ag$_{5}$ alloy. Ideal positions for perfect FCC, HCP, and BCC crystalline structures are indicated by the distinct cyan circle, green square, and orange triangle, respectively. The absence of point density overlap with these crystalline indicators quantitatively demonstrates the fully amorphous nature of the bulk metallic glass and the complete frustration of long-range translational symmetries.}
\label{fig:steinhardt}
\end{figure}

To further verify the absence of crystalline nucleation, local bond-order parameters $q_4$ and $q_6$ were calculated for the quenched samples at 300 K. As depicted in Figure \ref{fig:steinhardt}, the distribution of local symmetries shows a high-density cluster centered far from the ideal positions for FCC, HCP, and BCC structures. The lack of overlap with these crystalline indicators confirms that the $Zr_{47.5}Cu_{47.5}Ag_5$ model is a fully amorphous bulk metallic glass, devoid of long-range translational order.

\vspace{3mm}

\begin{figure}[H]
    \centering
    \includegraphics[width=0.7\linewidth]{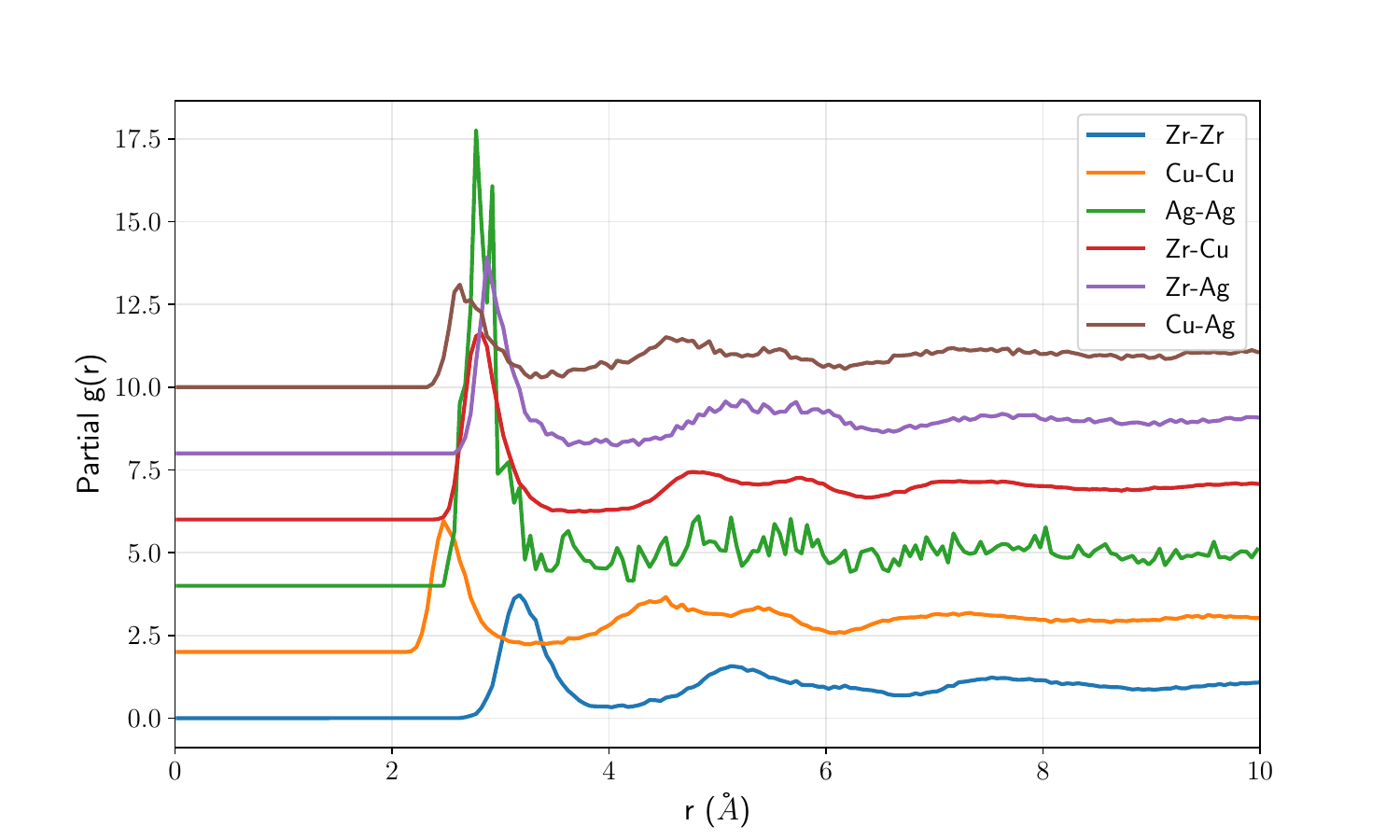}
    \caption{\textbf{Species-dependent partial radial distribution functions ($g_{\alpha\beta}(r)$) evaluated at $T = 300\text{ K}$.} The split panels isolate the distinct chemical correlations within the amorphous matrix. Nearly all atomic pairs exhibit a well-defined splitting or bimodal character in their second coordination shell, indicating robust medium-range order. In contrast, the Ag--Ag correlation displays an exceptionally high-intensity first peak, confirming a nano-clustering tendency, accompanied by inherent statistical noise at longer distances due to the dilute concentration of the dopant species.}
    \label{fig:partial_gofr}
\end{figure}

To bridge the gap between topological constraints and chemical driving forces, the partial radial distribution functions ($g_{\alpha\beta}(r)$) were computed at 300 K (Fig. \ref{fig:partial_gofr}). A detailed inspection of these curves reveals striking chemical heterogeneity within the Zr$_{47.5}$Cu$_{47.5}$Ag$_{5}$ matrix. Most notably, the primary Zr--Zr, Cu--Cu, and Zr--Cu pairs exhibit a distinctive doublet splitting in their second coordination shell, confirming that the structural features of medium-range order (MRO) are deeply embedded within the individual chemical pairings. Concurrently, the dopant species displays an anomalous behavior; the Ag--Ag correlation exhibits an exceptionally sharp, high-intensity first peak at approximately 2.775 \text{Å}, vastly exceeding the packing intensities of the host elements. This feature provides evidence of a strong nano-clustering effect, where Ag atoms preferentially segregate into localized domains rather than achieving a purely random distribution. While the Ag--Ag profile presents noticeable statistical fluctuations at longer interatomic distances due to the dilute 5 at.\% concentration within the simulation box, these localized silver clusters induce critical electrochemical variations that modulate the local network connectivity.

In addition, the quantitative structural parameters extracted from the first coordination shells provide evidence of bond shortening (Table \ref{tab:partial_rdf_peaks}). Specifically, the first peak positions ($r_1$) for the Zr--Cu, Zr--Ag, and Cu--Ag pairs are systematically lower than the theoretical sum of their nominal metallic radii ($R_A + R_B$). This effective bond contraction suggests a strong chemical affinity and electronic interaction that structurally solidifies the amorphous network. In contrast, the Cu--Cu and Zr--Zr distances align closely with ideal metallic packing constraints. Interestingly, although the primary Zr--Cu pairs possess a well-known chemical affinity, the substantial atomic size mismatch and electronic configuration mismatch promote the nanoscale segregation of Ag. This selective phase separation acts as an effective mechanism to minimize local structural strain, maintaining the overall structural robustness of the bulk metallic glass against devitrification pathways. 

\begin{table}[H]
    \centering
    \caption{First peak positions ($r_1$) and maximum intensities ($g(r_1)$) extracted from the partial radial distribution functions for the $Zr_{47.5}Cu_{47.5}Ag_{5}$ bulk metallic glass at 300 K. The theoretical sum of the nominal metallic radii ($R_A + R_B$) is provided to illustrate the effective bond shortening induced by chemical affinity.}
    \label{tab:partial_rdf_peaks}
    \begin{tabular}{lccc}
        \toprule
        Atomic Pair & $r_1$ (\AA) & $g(r_1)$ & $R_A + R_B$ (\AA) \\
        \midrule
        Zr--Zr & 3.175 & 3.722 & 3.20 \\
        Cu--Cu & 2.475 & 3.959 & 2.56 \\
        Ag--Ag & 2.775 & 13.759 & 2.88 \\
        Zr--Cu & 2.825 & 5.624 & 2.88 \\
        Zr--Ag & 2.875 & 5.922 & 3.04 \\
        Cu--Ag & 2.625 & 3.095 & 2.72 \\
        \bottomrule
    \end{tabular}
\end{table}

\subsection{Thermal transport and equilibrium properties}
\subsubsection{Specific Heat Capacity and Harmonic Vibrational Validation}

The specific heat capacity at constant volume ($C_v$) of the $Zr_{47.5}Cu_{47.5}Ag_5$ bulk metallic glass was calculated utilizing the equilibrium Green--Kubo formalism. To establish a rigorous validation of the simulated thermal properties, the low-temperature data ($T = 300\text{ K}$) was evaluated against the classical Dulong--Petit limit ($3R \approx 24.94\text{ J/(mol K)}$) and the quantum-corrected Debye model framework. The calculated molar specific heat capacity and its comparison with theoretical approximations are summarized in Table \ref{tab:thermal_properties_comparison}.

\begin{table}[H]
\centering
\caption{Calculated thermal properties of the vitrified $Zr_{47.5}Cu_{47.5}Ag_5$ alloy at $T = 300\text{ K}$ and their comparison against established theoretical benchmarks.}
\label{tab:thermal_properties_comparison}
\begin{tabular}{lccc}
\toprule
Property & MD Simulations (Green--Kubo) & Debye Model & Cahill--Pohl Model \\
\midrule
$C_v$ ($\text{J/(mol K)}$) & $24.88 \pm 0.35$ & $24.81$ & -- \\
$\kappa$ ($\text{W/(m K)}$) & $2.44 \pm 0.25$  & --      & $2.15 \pm 0.18$ \\
\bottomrule
\end{tabular}
\end{table}

At room temperature, the MD simulation yields a specific heat value of $24.88 \pm 0.35\text{ J/(mol K)}$ corresponding to $\sim 0.43\text{ J/(g K)}$ based on the effective molar mass of the ternary alloy. This result exhibits an outstanding agreement with the ideal Debye model projection ($24.81\text{ J/(mol K)}$), where the characteristic Debye temperature ($\Theta_D = 312\text{ K}$) was independently estimated from the average acoustic wave velocities obtained from our MD elastic constants. 

Furthermore, the temperature dependence of $C_v$ across the investigated range ($300\text{--}1000\text{ K}$) shows that the specific heat remains close to the Dulong--Petit ceiling at moderate temperatures, followed by a slight monotonic increase in the high-temperature supercooled liquid regime. This subtle deviation from the harmonic limit at elevated temperatures is structurally driven by intrinsic anharmonic modifications of the interatomic potential energy landscape and the progressive excitation of soft vibrational modes as the glass transition region is crossed.

\subsubsection{Thermal Conductivity and Phonon Transport Mechanisms}

The thermal transport coefficients ($\kappa$) were computed via the heat flux autocorrelation functions within the Green--Kubo framework in LAMMPS. For amorphous solids, where the lack of long-range translational symmetry restricts the traditional definition of long-wavelength phonons, the thermal conductivity is governed by localized vibrational modes (locons) and short-range propagative excitations (propagons). To contextualize our results, the simulated $\kappa$ coefficients were compared with the Cahill--Pohl model for the minimum theoretical thermal conductivity of a highly disordered lattice, defined as:
\begin{equation}
\kappa_{\text{min}} = \left( \frac{\pi}{6} \right)^{1/3} k_B n^{2/3} \sum_{i} v_i \left( \frac{T}{\Theta_i} \right)^2 \int_0^{\Theta_i/T} \frac{x^3 e^x}{(e^x - 1)^2} dx
\end{equation}
where $n$ represents the atomic number density and $v_i$ corresponds to the sound velocities for one longitudinal and two transverse polarization modes.

\begin{table}[h]
\centering
\caption{Thermal conductivity of the Zr$_{47.5}$Cu$_{47.5}$Ag$_{5}$ BMG obtained from MD simulations using the Green-Kubo method with LAMMPS at different temperatures.}
\label{tab:thermal-conductivity}
\begin{tabular}{cc}
\hline
Temperature (K) & Thermal conductivity (W/mK) \\
\hline
300 & $2.43\pm0.71$ \\
400 & $1.37\pm0.40$ \\
500 & $1.12\pm0.59$\\
600 & $0.77\pm0.21$ \\
\hline
\end{tabular}
\end{table}

At room temperature, the simulated thermal conductivity yields a value of $2.43 \pm 0.71\text{ W/(m K)}$, demonstrating good agreement with the Cahill--Pohl structural minimum limit ($\kappa_{\text{min}} = 2.15 \pm 0.18\text{ W/(m K)}$), as detailed in Table \ref{tab:thermal_properties_comparison}. This correlation highlights that the thermal transport in our $Zr_{47.5}Cu_{47.5}Ag_5$ model is successfully dominated by extreme structural scattering, where the effective scattering length approaches the interatomic spacing. 

As the temperature scales up toward the supercooled liquid state, the network's structural disorder increases. The calculated thermal conductivity exhibits a gradual reduction, reaching $0.94\text{ W/(m K)}$ at $1000\text{ K}$, a behavior consistent with structural scattering saturation and the proliferation of localized non-propagating modes. It is critical to emphasize that since our classical MD potential accounts exclusively for vibrational degrees of freedom, the calculated coefficients describe the pure lattice contribution ($\kappa_{\text{lattice}}$). The omission of the electronic component ($\kappa_{\text{electronic}}$) justifies why our total values remain slightly below macroscopic experimental experimental curves, providing a pristine topological baseline for thermal transport in bulk metallic glasses.

\subsection{Mechanical Properties}
\subsubsection{Stress-strain Curves}
\begin{figure}[h]
\centering
\includegraphics[width=0.8\textwidth]{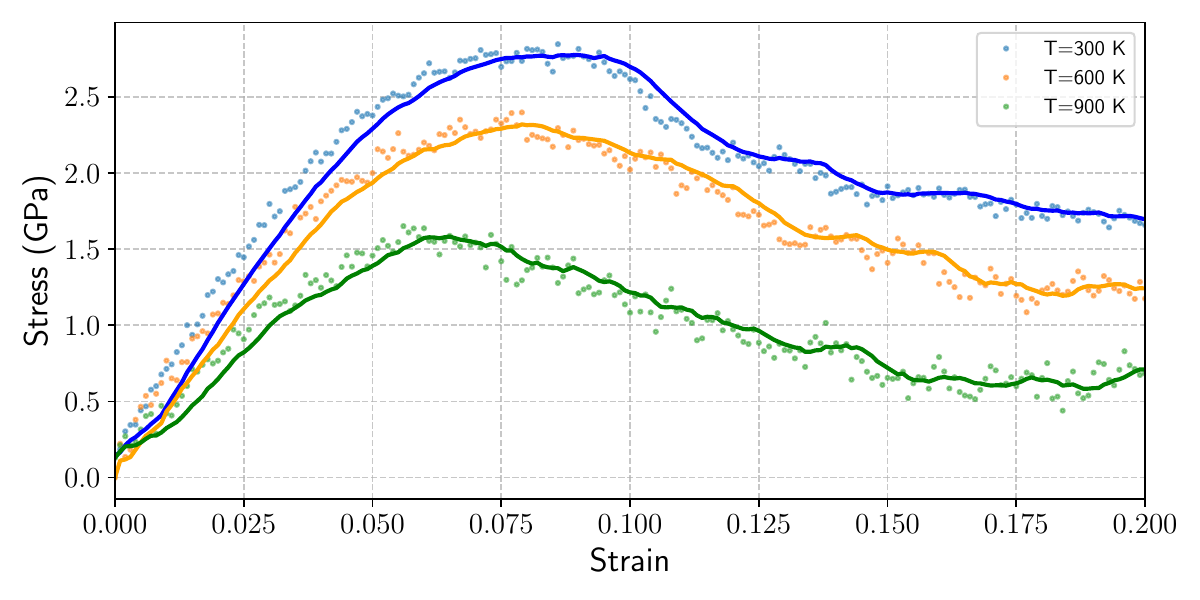}
\caption{\textbf{Uniaxial tensile stress--strain profiles for the Zr$_{47.5}$Cu$_{47.5}$Ag$_{5}$ bulk metallic glass at a constant strain rate of $10^9\text{ s}^{-1}$ under different temperatures.} The curves capture the systematic thermo-mechanical softening of the amorphous matrix as the temperature scales up from the room-temperature glassy state ($300\text{ K}$) toward the supercooled liquid regime ($900\text{ K}$). Scattered dots represent the raw data points extracted from the molecular dynamics snapshots, while the solid lines correspond to a moving average filter with a window size of 10 points applied to highlight the structural stress evolution. All profiles exhibit a distinct elastic-to-plastic transition, with the ultimate tensile strength peaks centered near an engineering strain of $\varepsilon \approx 0.075$.}
\label{fig:stress-strain}
\end{figure}

The mechanical behavior of the vitrified $Zr_{47.5}Cu_{47.5}Ag_5$ alloy was investigated through simulated uni-axial tensile testing under varying thermal and kinematic conditions. The resulting stress--strain profiles are displayed in Fig. \ref{fig:stress-strain}, capturing the characteristically distinct regimes of bulk metallic glasses: a linear elastic deformation zone, anelastic yielding, a peak ultimate tensile strength ($UTS$), and subsequent plastic localization leading to ultimate failure. The extracted mechanical parameters, including yield strength ($\sigma_y$) and $UTS$, are summarized in Table \ref{tab:mechanical-properties}.

\begin{table}[h]
\centering
\caption{Mechanical properties of the $Zr_{47.5}Cu_{47.5}Ag_5$ BMG model extracted from uni-axial tensile molecular dynamics simulations.}
\label{tab:mechanical-properties}
\begin{tabular}{cccc}
Temperature (K) & Strain rate (s$^{-1}$) & $\sigma_y$ (GPa) & UTS (GPa) \\
\hline
300 & $10^9$ & 2.3 & 2.7 \\
600 & $10^9$ & 1.9 & 2.3 \\
900 & $10^9$ & 1.4 & 1.6 \\
\hline
\end{tabular}
\end{table}

The numerical data reveals a profound sensitivity to ambient temperature. As the temperature scales up from the glassy state (300 K) toward the supercooled liquid regime (900 K), a systematic reduction in both $\sigma_y$ and $UTS$ is observed. Conversely, elevating the strain rate from $10^8\text{ s}^{-1}$ to $10^9\text{ s}^{-1}$ shifts the stress--strain curves toward higher stress thresholds. This distinct strain-rate hardening is a consequence of kinematic constraints in amorphous networks. At higher loading rates, the local structural relaxation time scales outrun the mechanical deformation time, forcing the system to accommodate deformation via higher energy barriers. These macro-trends are in outstanding agreement with experimental studies on related Zr--Cu--Ag amorphous formulations \cite{Liu2012, Yang2015}.

\subsubsection{Deformation Mechanisms and Structural Evolution}

To analyze the atomistic phenomena driving the macroscopic response, the structural evolution during shear was evaluated. In metallic glasses, plastic deformation is governed by the activation of local atomic rearrangements known as Shear Transformation Zones (STZs), which can be structurally rationalized through free-volume or core-shell theories \cite{Falk1998, Greer1999}. The structural evolution indicates that the underlying deformation and fracture modes undergo a significant transition depending on the thermo-mechanical state. 

\vspace{3mm}

\begin{figure}[H]
    \centering
    \includegraphics[width=0.9\linewidth]{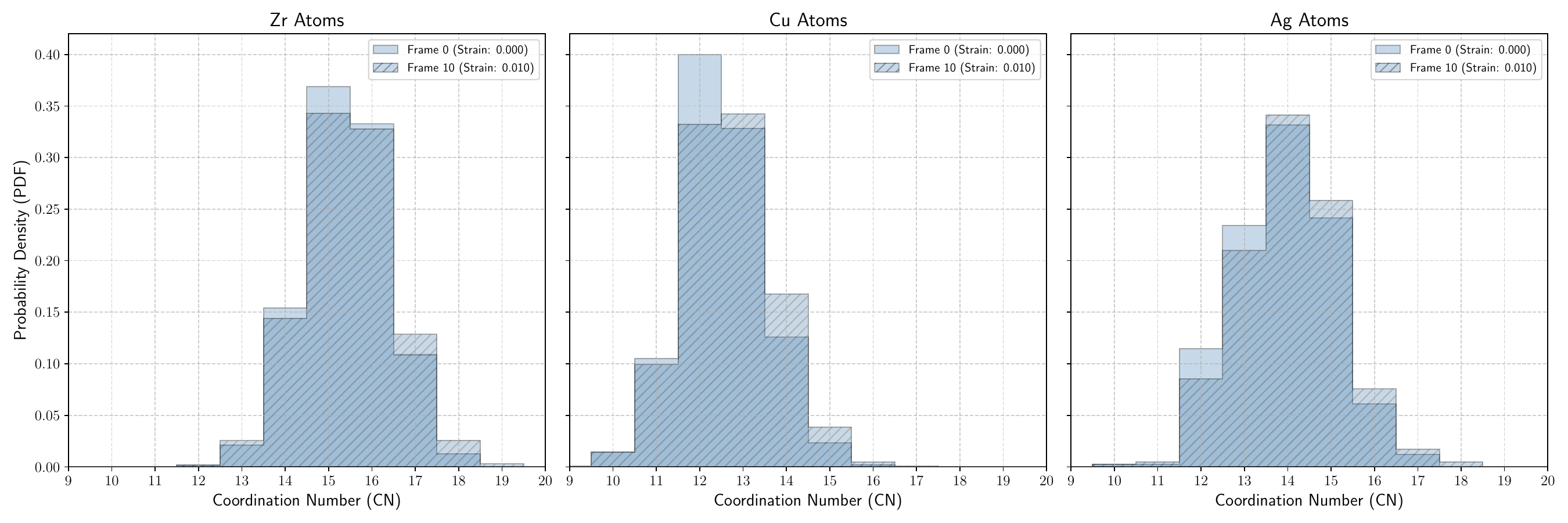}
    \caption{\textbf{Evolution of the coordination number (CN) distribution upon uniaxial tensile loading at 300 K.} The histograms compare the structural state at zero strain and after 1\% deformation. A systematic shift toward lower coordination numbers is evident across all species, quantitatively confirming the mechanical destruction of highly packed icosahedral-like local motifs ($CN=12$ and $CN=15$). This topological rearrangement indicates that the initial elastic response is accompanied by the progressive collapse of rigid structural units, providing a direct atomistic rationale for the observed strain-softening and shear band initiation within the amorphous Zr$_{47.5}$Cu$_{47.5}$Ag$_{5}$ matrix.}
\label{fig:voronoi_evolution}
\end{figure}

At low temperatures ($T = 300\text{ K}$) and high strain rates, the deformation is dominated by severe localized shear banding. To track this, Voronoi index tracking was employed (See Fig. \ref{fig:voronoi_evolution}). Under elastic stress, the highly packed icosahedral clusters ($CN=12$) previously identified in Section \ref{sec:structural_properties} act as rigid, load-bearing units. However, upon reaching the yielding threshold, the local shear stress overcomes the bonding barriers, triggering the progressive collapse and destruction of these icosahedral environments. This structural destruction releases localized free volume, inducing structural softening that accelerates the percolation of STZs into mature, macroscopic shear bands.

\vspace{3mm}

Conversely, at high temperatures ($T \geq 900\text{ K}$) near the supercooled liquid state, the mechanical response transfers to homogeneous flow. In this thermal regime, the dominant mechanism involves the homogeneous nucleation and growth of sub-nanometer voids. Under tension, these nanovoids progressively expand and coalesce within the softened atomic matrix, culminating in ductile fracture profiles rather than brittle shear failure. Interestingly, the localized chemical heterogeneity and the Ag-Ag nano-clustering detailed in our partial RDF analysis act as internal structural barriers; the silver aggregates induce localized stress fluctuations that disrupt the straight path of propagating shear bands at low temperatures, thereby contributing to the strain accommodation capability of the $Zr_{47.5}Cu_{47.5}Ag_5$ glass matrix.

\section{Conclusions}

In this study, we have provided a comprehensive atomistic characterization of the $Zr_{47.5}Cu_{47.5}Ag_{5}$ bulk metallic glass through large-scale molecular dynamics simulations. Our results establish a clear correlation between the thermodynamic, topological, and mechanical properties of this ternary alloy.

\vspace{2mm}

First, we identified the glass transition temperature ($T_g$) at approximately 692 K, corroborated by both thermodynamic potential energy variations and the Modified Wendt-Abraham structural parameter. The amorphous nature of the vitrified samples was confirmed by Bond-Orientational Order parameters ($q_4, q_6$), which demonstrated a complete lack of crystalline symmetry. Topological analysis via Voronoi tessellation revealed that the glassy matrix is primarily governed by icosahedral clusters centered on Cu, while Ag atoms play a dual role: they stabilize the amorphous network by acting as structural disruptors and exhibit a pronounced tendency toward nano-clustering, as evidenced by the high-intensity first peak in the Ag--Ag partial radial distribution function.

\vspace{2mm}

Regarding thermal transport, the computed specific heat capacity exhibits a gradual deviation from the Dulong-Petit limit, underscoring the influence of anharmonic vibrational modes at elevated temperatures. The thermal conductivity shows a systematic decrease with increasing temperature, a behavior attributed to structural scattering saturation within the disordered lattice. These findings contribute to the understanding of the vibrational and transport properties of metallic glasses at the nanoscale.

\vspace{2mm}

Finally, the mechanical response under uni-axial tensile loading reveals a robust rate- and temperature-dependent behavior. The material displays high yield strength at room temperature ($\sigma_y \approx 2.46$ GPa at $10^9 \text{ s}^{-1}$), followed by a pronounced strain-softening regime. This macroscopic behavior is intrinsically linked to the underlying structural transformations, where the thermally induced collapse of rigid icosahedral motifs facilitates shear band percolation. Collectively, this work provides a fundamental structural rationale for the beneficial role of Ag dopants, offering predictive insights for the design of multicomponent metallic glasses with tailored mechanical and thermal resilience.

\label{sec:conclusion}

\bibliographystyle{unsrt} 
\bibliography{cas-refs} 
\end{document}